\documentclass[12pt]{article}
\newcommand{\be}{\begin{eqnarray}}
\newcommand{\en}{\end{eqnarray}}

\begin{document}
\begin{titlepage}
\setlength{\textwidth}{5.9in}
\begin{flushright}
EFI 2000-51\\

\end{flushright}
\begin{center}
\vskip 0.3truein
{\Large\bf {Reduction of Coupling Parameters}} \\
\vskip 0.15truein
{\Large\bf {in Quantum Field Theories}} 
\footnote{For the {\it `Concise Encyclopedia of SUPERSYMMETRY'}, \\
 Kluwer Academic Publishers, Dortrecht, (Editors: Jon Bagger, Steven Duplij
and Warren Siegel) 2001.}

\vskip0.4truein
{Reinhard Oehme}
\footnote{E-mail: oehme@theory.uchicago.edu}
\vskip0.4truein
{\it Enrico Fermi Institute and Department of Physics}\\
{\it University of Chicago} \\
{\it Chicago, Illinois, 60637, USA}
\end{center}

\vskip0.4truein

\centerline{\bf Abstract}

\vskip0.2truein
A concise survey is given of the general method of reduction 
in the number of coupling parameters. Theories with several 
independent couplings are related to a set of theories with
a single coupling. The reduced theories may or may not have particular
symmetries. A few have asymptotic power series expansions, others 
contain non-integer powers and/or logarithmic factors. 
An example is given with two power series solutions, one with
N = 2 Supersymmetry, and one with no known symmetry. In a second
example, the reduced Yukawa coupling of the superpotential in a
dual magnetic supersymmetric  gauge theory is uniquely given by
the square of the magnetic gauge coupling with a known factor. 

\end{titlepage}
\newpage
\baselineskip 18 pt
\pagestyle{plain}
\setlength{\textwidth}{5.9in}

\noindent REDUCTION OF COUPLING PARAMETERS
in multi parameter quantum field theories [1,2] often leads
to one-parameter theories with supersymmetry or
other symmetries. In addition, there may be solutions
with no recognizable symmetry. For pairs of dual
supersymmetric gauge theories, which require the presence
of superpotentials, the reduction method makes it possible
to obtain dual pairs with each theory having a single 
coupling parameter[3]. There are many other applications 
of the reduction Method [4,5].

\noindent Consider quantum field theories with several
dimensionless coupling parameters

$\lambda,\lambda_1,\ldots,\lambda_n$.

\noindent The corresponding effective couplings
$\overline{\lambda}(u),\overline{\lambda}_k(u), k=1,...,n$
\noindent are functions of scaling parameter $u=k^2/\kappa^2$, where
$\kappa^2<0$ is the normalization point. They satisfy the
renormalization group equations

$u{(d \overline{\lambda}}/{d u})
= \beta(\overline{\lambda},\overline{\lambda}_1,
\dots,\overline{\lambda}_n),~~~
u{(d \overline{\lambda}_k}/{d u})
=\beta_k(\overline{\lambda},\overline{\lambda}_1,
\dots,\overline{\lambda}_n)~$,

\noindent where the functions $\beta$ and $\beta_k$ are obtained
from the vertex terms of the theory. With
$\overline{\lambda}(u)$ being an analytic function,
one can choose a point where
$(d \overline{\lambda}(u)/du)\neq 0$ and introduce $\overline{\lambda}(u)$
as a new variable in these equations.
With $\overline{\lambda}_k(u)\rightarrow
\lambda_k(\overline{\lambda}(u))$, and
$\overline{\lambda}(u)\rightarrow\lambda$, this substitution
eliminates the variable $u$ and yields the
REDUCTION EQUATIONS
\begin{eqnarray}
\beta(\lambda) \frac{d\lambda_k (\lambda)}{d\lambda} = \beta_k (\lambda)~,
~~~k=1,\ldots,n~.
\end{eqnarray}
Here $\beta(\lambda)=\beta(\lambda,\lambda_1,\ldots,\lambda_n)$
and $\beta_k(\lambda)=\beta_k(\lambda,\lambda_1,\ldots,\lambda_n)$,
with the insertions $\lambda_k = \lambda_k (\lambda),
~k = 1,\ldots,n ~$.
The reduction equations are necessary and sufficient for
the Green's functions of the one-parameter theory

$G(k_i,\kappa^2,\lambda) = G \left( k_i, \kappa^2, \lambda,
\lambda_1 (\lambda),\ldots,\lambda_n (\lambda)\right)$

\noindent to satisfy the renormalization group equations in the single variable
$\lambda$:
\begin{eqnarray}
\left(\kappa^2 \frac{\partial}{\partial \kappa^2} + \beta(\lambda)
\frac{\partial}{\partial\lambda} + \gamma_G(\lambda) \right)
G(k_i,\kappa^2,\lambda) ~ = ~ 0 ~~,
\end{eqnarray}
where $\beta(\lambda)$ and $\gamma_{G}(\lambda)$ are
again given by the
corresponding coefficients of the multi-parameter theory
with insertions.
Of course, one can also require the validity of eq.(2)
in order to obtain the reduction equations (1).
A priori, the reduction scheme
is very general. But for most applications considered, the functions
$\lambda_k (\lambda)/\lambda = f_k(\lambda)$ are bounded for
$\lambda \rightarrow 0$. Furthermore, the $\beta$-functions are represented
by asymptotic power series in the weak coupling limit:
\begin{eqnarray}
\beta(\lambda,\lambda_1,\dots,\lambda_n)
=\beta_{0}\lambda^2+
(~\beta_{1}\lambda^3+
\beta_{1k}\lambda_k\lambda^2+
\beta_{1k k'}\lambda_k\lambda_{k'}\lambda~) + \cdots ,\\
\beta_{k}(\lambda,\lambda_1,\dots,\lambda_n)
=(c_{k}^{(0)}\lambda^2+
c_{k, k'}^{(0)}\lambda_{k'}\lambda+
c_{k, k' k''}^{(0)}\lambda_{k'}\lambda_{k''}~)+ \cdots.
\end{eqnarray}
\noindent It is seen that the the reduction equations are singular at the
origin. This implies that the Picard-Lindeloef theorem about the
uniqueness of solutions does not apply. Using equivalence
transformations, possible mass and gauge parameter dependencies
of the coefficient functions can be removed.
With the original $\beta$-functions given as asymptotic
power series expansions, solutions $\lambda_k(\lambda)$
of the reduction equations are considered which are
also of the form of asymptotic expansions.
Of special interest are solutions in the form of power series
expansions, but in general, non-integer
powers as well as logarithmic terms are possible.
Consider first power series solutions
\begin{eqnarray}
\lambda_k(\lambda)=\lambda f_k(\lambda),~~~~
f_k(\lambda)=f_{k}^{0}+\sum_{m=1}^{\infty}\chi_{k}^{(m)}\lambda^m .
\end{eqnarray}
\noindent Substitution into the reduction equations
yields the fundamental one-loop relation
\begin{eqnarray}
c_{k}^{(0)}+
(c_{kk'}^{(0)}-\beta_0\delta_{k k'})f_{k'}^{0}+
c_{kk'k''}^{(0)}f_{k'}^{0}f_{k''}^{0}=0~.
\end{eqnarray}
Given a solution $f_k^0$ of these quadratic equations,
the one-loop criteria
\begin{eqnarray}
det \left( M_{kk^\prime}(f^0)- m \beta_{0}\delta_{kk^\prime}\right)
\not = 0
~~ for ~~m=1,2,\ldots,
\end{eqnarray}
\begin{eqnarray}
M_{kk^\prime} (f^0) = c^{(0)}_{k,k^\prime} + 2 c^{(0)}_{k,k^\prime
k^{\prime\prime}} f^0_{k^{\prime\prime}} - \delta_{kk^\prime}
\beta_{0}~.
\end{eqnarray}
\noindent are sufficient to insure that all coefficients
$\chi^{(m)}$ in the expansion of $f_k(\lambda)$ are determined.
Then the reduced
theory has a power series expansion in $\lambda$, and
all possible solutions of this kind are determined by the
one-loop equation for $f^0_k$.
With the coefficients $\chi^{(m)}$ fixed, one can use regular
reparametrization transformations in order to remove all but
the first term in the expansion of the
functions $f_k(\lambda)$. These reparametrization transformations
are of the form
\noindent

$\lambda^\prime=\lambda^\prime (\lambda,\lambda_1,\ldots,\lambda_n) =
\lambda + a^{(20)}
\lambda^2 + a_k^{(11)} \lambda_k \lambda + \cdots~$,

$\lambda^\prime_k=\lambda^\prime_k
(\lambda,\lambda_1,\ldots,\lambda_n) = \lambda_k +
b^{(20)}_{kk^\prime k^{\prime\prime}} \lambda_{k^\prime}
\lambda_{k^{\prime\prime}}
+ b^{(11)}_{kk^\prime} \lambda_{k^\prime} \lambda + \cdots ~$.

\noindent They leave one-loop quantities invariant.
Given the validity of the conditions (7), there is then a frame
where the solutions are of the form
\begin{eqnarray}
\lambda_k (\lambda) = \lambda f^0_k ~ ,
\end{eqnarray}
with the coefficients $f^0_k$ determined by the one-loop reduction
equations (6). These usually have only a few characteristic solutions.
In the special case where $f^0_k=0$, and $\chi_k^{(m)}=0$ for
$m<N$, one has $f_k(\lambda) = \chi_k^{(N)}\lambda^N $ after
an appropriate reparametrization.

\noindent Besides the power series solutions, reduced to the form (9),
there can be `general' solutions of eqs.(1) which approach the same
limit $f_k^0$, but contain non-integer powers. For example,
if the matrix $\beta_0^{-1}M(f^0)$ has one non-integer eigenvalue $\eta>0$,
then there is a solution
$f_k(\lambda) = f_k^0 + {\chi}_k^{(\eta)} \lambda^{\eta}+ \cdots$,
after reparametrization. The coefficient $\chi_k^{(\eta)}$
contains r free parameters if the eigenvalue has r-fold degeneracy.
All other coefficients are determined.
\noindent Questions about the stability of solutions have been
discussed in connection with the Lyapunov-Malkin theorems [3].

\noindent In case the determinant (7) should vanish due to some
positive eigenvalue $m=N$, then the asymptotic series solution
contains in geneal terms like $\lambda^N log\lambda$.

\vskip0.1truein

The essential features of the reduction method are best
seen by evaluating cases of particular interest.
For a reduction resulting in SUSY and
non-SUSY theories, one can consider a gauge theory
with one Dirac field, one scalar and one pseudoscalar field,
all in the adjoint representation of SU(2) [1,2].
Besides the usual gauge couplings, the direct interaction
part of the Lagrangian is given by
\begin{eqnarray}
{L}_{dir.int.} &=& i\sqrt{\lambda_1}~
\epsilon^{abc} \overline{\psi}^a (A^b + i\gamma_5 B^b ) \psi^c \cr
&-& \frac14 \lambda_2 (A^a A^a + B^a B^a)^2 +
\frac{1}{4} \lambda_3 (A^a A^b + B^a B^b)^2 ~~.
\end{eqnarray}
Writing $\lambda=g^2$, where $g$ is the gauge coupling, and
$\lambda_k=\lambda f_k$,
with k=1,2,3 ,
the one-loop $\beta$-function coefficients of this theory are

\noindent
$(16\pi^2)\beta_{g0}=-4,~
(16\pi^2)\beta_1^{0}=8f_1^2-12f_1,~
(16\pi^2)\beta_2^{0}=3f_3^2-12f_3f_2+14f_2^2+8f_1f_2-8f_1^2-12f_2+3,~
(16\pi^2)\beta_3^{0}=-9f_3^2+12f_3f_2+8f_3f_1-12f_3-3. $

\noindent The algebraic reduction equations (4) have four real solutions:

$(f^0_1=1,~f^0_2=1,~f^0_3=1),~~~
(f^0_1=1,~f^0_2=\frac{9}{\sqrt{105}},~f^0_3=\frac{7}{\sqrt{105}})$,

\noindent and two others with reversed signs of $f^0_2$ and $f^0_3$,
so that the classical potential approaches negative infinity
with increasing magnitude of the scalar fields. These latter
solutions will not be considered further. 

\noindent The eigenvalues of the matrix $\beta_{g0}^{-1}M(f^0)$
are respectively $\left(-2, -3, +\frac{1}{2}\right)$ and 

\noindent
$\left(-2, -\frac{3}{4}~\frac{25+\sqrt{343}}{\sqrt{105}}, -\frac{3}{4}~\frac
{25-\sqrt{343}}{\sqrt{105}}\right)
=(-2, -3.189..., -0.470...).$

\noindent There are no positive integers appearing in these expressions.
Hence the coefficients of the power series solutions are
determined and can be removed by reparametrization, except
for the invariant first term.
Then the solutions are
\begin{eqnarray}
(a) ~~~\lambda_1 = \lambda_2 = \lambda_3 = g^2 ~,
\end{eqnarray}
which corresponds to an $N = 2 $ extended SUSY Yang-Mills theory, and
\begin{eqnarray}
(b) ~~~\lambda_1 = g^2,~~ \lambda_2 = \frac{9}{\sqrt{105}}~ g^2, ~~
\lambda_3 = \frac{7}{\sqrt{105}}~ g^2 ~,
\end{eqnarray}
which is not associated with any known symmetry, at least
in four dimensions. Both theories are `minimally' coupled gauge theories
with matter fields. The eigenvalues
of the matrix $\beta_{g0}^{-1}M(f^0)$ are all negative with the
exception of the third one for the N=2 supersymmetric theory. In this case
there exists a general solution with
$\eta=+\frac{1}{2}$, and with the coefficient given
by $\chi^{(\frac{1}{2})}=(0,C,3C)$, where $C$ is an arbitrary parameter. The
theory with $C\ne0$ corresponds to one with hard breaking of SUSY. It has an
asymptotic power series in
$g$ and not in $g^2$, as is the case for the invariant theory.

\noindent From the present example, and many others, one realizes that the
special frame, where the power series solutions of the reduction
equations are of the simple form (9), is
a natural frame as far as the reduced one-parameter theories are concerned.
The $\beta$-functions of the reduced theories are still power series and are
not reduced to polynomials.

\vskip0.1truein

As another application of the reduction method, the "magnetic" gauge
theory is considered which is the dual of SQCD as the "electric" theory [3].
For SQCD the gauge group is $SU(N_C)$ with $N=1$ SUSY, and there
are $N_F$ quark superfields in the fundamental representation. There is
only one coupling parameter, the gauge coupling $g_e$ . The
$\beta$-function is given by the asymptotic expansion

$\beta_e (g_e^2)=\beta_{e0}g_e^4+\beta_{e1}g_e^6+ \cdots$,

\noindent with one loop coefficient

$\beta_{e0}=(16\pi^2)^{-1} (-3N_C + N_F )$.

\noindent It is proposed that there exists a dual
magnetic theory which provides an alternate description
at low energies [6]. But both theories coincide only at the non-trivial
infrared fixed point in the conformal window
$\frac{2}{3}N_C<N_F<3N_C$.
For appropriate
values of $N_C$ and $N_F$,
the magnetic theory has the gauge group $G^d=SU(N^d_C)$
with $N^d_C=N_F-N_C$.
There are $N^d_F=N_F$ quark superfields $q$ in the fundamental
representation, the corresponding
anti-quark superfields $\overline{q}$, and $N_F^2$ independent
scalar superfields $M$, which are coupled via a Yukawa superpotential
of the form $\sqrt{\lambda}M^i_jq_i{\overline{q}}^j$.
This coupling is required by the anomaly matching conditions, which are
used in the construction of dual theories, and by the need for both
theories to have the same physical symmetries. In the conformal window,
the potential drives the magnetic theory to the infrared fixed point. The
$\beta$-functions of the magnetic theory can be written in the form

$\beta_m (g_m^2, \lambda)=\beta_{m0}~g_m^4~+~(\beta_{m1}~g_m^6~+~
\beta_{m1,\lambda}~g_m^4\lambda) ~+~\cdots $

$\beta_\lambda (g_m^2, \lambda)~=~c_\lambda g_m^2 \lambda ~+~
c_{\lambda\lambda} \lambda^2 ~+~\cdots ~$ ,

\noindent with the relevant lowest order coefficients given by

$\beta_{m0}=(16\pi^2)^{-1} (3N_C - 2N_F )$

$c_\lambda=(16\pi^2)^{-1} \left( -4 \frac{(N_F - N_C)^2 - 1}{2(N_F -
N_C)} \right)$

$c_{\lambda\lambda}=(16\pi^2)^{-1} \left(3N_F - N_C) \right)$.

\noindent
At first, the reduction method is applied to the magnetic
theory in the conformal window $\frac{2}{3}N_C<N_F<3N_C$.
There are two power series solutions. After reparametrization,
one solution is given by [3]
\begin{eqnarray}
\lambda_1(g_m^2) = g_m^2 f(N_C, N_F), 
\end{eqnarray}
\begin{eqnarray}
f(N_C, N_F)=\frac{\beta_{m0} - c_{\lambda}}{c_{\lambda\lambda}}=
\frac{N_C\left( N_F - N_C - {2}/{N_C} \right)}
{(N_F - N_C)(3N_F - N_C)}.  
\end{eqnarray}

\noindent The other solution is
$\lambda_2(g_m^2) \equiv 0$. Since the latter removes the superpotential, it
is excluded, and one is left with a unique single power series solution.
This solution implies a theory with a single gauge coupling $g_m$,
and renormalized perturbation expansions which are power series in
$g_m^2$. It is the appropriate dual of SQCD. There are `general' solutions,
but they all approach the excluded power solution $\lambda_2(g_m^2) \equiv
0$.
With one exception, they involve non-integer powers of $g_m^2$.
The reduction can be extended to the `free electric region' $N_F > 3N_C$,
and to the `free magnetic region' $N_C+2 < N_F < \frac{2}{3}N_C$, which is
non-empty for $N_C>4$.
The results are similar.
In the free magnetic case, one deals however with the approach to a trivial
infrared fixed-point.

\noindent Further applications of the reduction method in connection with
duality may be found in [4].
Dual theories can be obtained as appropriate
limits of brane systems. In these brane constructions,
duality corresponds essentially to a reparametrization
of the quantum moduli space of vacua
of a given brane structure. It remains
to find out how the reduction solutions are related to special
features of these constructions. 

\noindent There are also more phenomenological uses for the reduction
schemes, in particular within the framework of
supersymmetric grand unified theories [5].

\newpage

\noindent BIBLIOGRAPHY

[1] R. Oehme and W. Zimmermann,
MPI Report MPT-PAE/Pth 60/82 (1982),
Commun. Math. Phys. {\bf 97} (1985) 569;
R. Oehme, K. Sibold and W. Zimmermann,
Phys. Lett. {\bf B147} (1984) 115; Phys. Lett. {\bf B153} (1985) 142.

[2] W. Zimmermann,
Commun. Math. Phys. {\bf 97} (1985) 211
(Symanzik Memorial Volume);
R. Oehme,
CERN-Report TH.42-45/1985,
Prog. Theor. Phys. Suppl. {\bf 86} (1986) 215
(Nambu Festschrift);
K. Sibold,
Acta Physica Polonica, {\bf 19} (1988) 295.
(These papers contain reference to earlier work on
reduction methods)

[3] R. Oehme, Phys. Lett. {\bf B399} (1997) 67; 
Phys. Rev. {\bf D59} (1999) 105004; 
R. Oehme, `Reduction of Coupling Parameters and Duality'
in {\it Recent Developments in Quantum Field Theory}, edited by
P. Breitenlohner, D. Maison and J. Wess
(Springer Verlag, Heidelberg, New York, 2000), hep-th/9903092.

[4] E. Gardi and G. Grunberg, JHEP {\bf 024} (1999) 9903.

[5] T. Kobayashi, J. Kubo, M. Mondragon and G. Zoupanos,
Nucl. Phys.  {\bf B511} (1998) 45;
J. Kubo, `Applications of Reduction of Couplings',
in {\it Recent Developments in Quantum Field Theory}, edited by
P. Breitenlohner, D. Maison and J. Wess
(Springer Verlag, Heidelberg, New York, 2000), hep-th/9903482, 
(This article contains detailed references to phenomenological
applications).

[6] N. Seiberg, Phys. Rev. {\bf D49} (1994) 6857; Nucl. Phys. {\bf B435} (1996) 129.

\vskip0.1truein

\noindent {\it Reinhard Oehme}

\end{document}